\documentclass[prl,twocolumn,superscriptaddress]{revtex4}

\usepackage{tabularx}
\usepackage{float}
\usepackage{bm}
\usepackage{graphicx}
\usepackage{amsfonts}
\usepackage{amsmath}
\usepackage{mathtools}
\usepackage{color}
\usepackage{dsfont}
\usepackage{mathrsfs}
\usepackage{hyperref}
\usepackage{breqn}

\usepackage{subfiles}

\makeatletter
\let\cat@comma@active\@empty
\makeatother

\begin{document}

\title{Chiral waveguide optomechanics: first order quantum phase transitions with $\mathbb{Z}_3$ symmetry breaking.}

\author{D. D. Sedov}
\affiliation{Department of Physics and Engineering, ITMO University, Saint Petersburg 197101, Russia}

\author{V. K. Kozin}
\affiliation{Science Institute, University of Iceland, Dunhagi-3, IS-107 Reykjavik, Iceland}
\affiliation{Department of Physics and Engineering, ITMO University, Saint Petersburg 197101, Russia}

\author{I. V. Iorsh}
\affiliation{Department of Physics and Engineering, ITMO University, Saint Petersburg 197101, Russia}

\begin{abstract}
We present a direct mapping between the quantum optomechanical problem of the atoms harmonically trapped in the vicinity of a chiral waveguide and a generalized quantum Rabi model and discuss the analogy between the self-organization of atomic chains in photonic structures and Dicke-like quantum phase transitions in the ultrastrong coupling regime. We extend the class of the superradiant phase transitions for the systems possessing $\mathbb{Z}_3$ rather than parity $\mathbb{Z}_2$ symmetry and demonstrate the emergence of the multicomponent Schrodinger cat ground states in these systems.
\end{abstract}
\maketitle
The arrays of quantum emitters coupled to a common one-dimensional photonic reservoir are the main object studied by the emerging field of waveguide quantum electrodynamics (WQED)~\cite{Roy2017,KimbleRMP2018}. The field currently experiences a rapid progress due to the tremendous developments in quantum technologies allowing realizations of this type of systems based on a variety of platforms including superconducting qubits~\cite{vanLoo2013,Mirhosseini2019}, cold-atoms~\cite{Corzo2019} or semiconductor quantum dots~\cite{foster2019tunable}. The key features of waveguide quantum optical set-ups are the emergent long-range correlations between the qubits harnessed through the exchange of the propagating waveguide photons, and the inherent open nature of these systems, provided by the leakage of the photons. Recently, the set-ups comprising the ring-shaped topological waveguides have been suggested~\cite{PhysRevB.101.205303, mehrabad2019chiral} which combine the long range inter-qubit correlations and quasi-hermiticity. These set-ups could be particularly useful for the emulation of the strongly correlated quantum models, since the latter are usually Hermitian ones.

One of the factors, limiting the diversity of the quantum many-body phenomena supported by the WQED set-ups is the relatively small radiative coupling of the individual qubits to the photonic mode as compared to the transition frequencies. This leaves us in the weak coupling region of the light-matter interaction. At the same time, reaching the regime of the ultrastrong coupling~\cite{FriskKockum2019,RevModPhys2019} at which the coupling strength becomes comparable with the transition frequencies would enable the access to a plethora of fascinating quantum phenomena, such as non-vacuum and correlated ground states, and possible application in quantum memory~\cite{kyaw2015scalable} and quantum metrology~\cite{colley2007mid,ruggenthaler2018quantum}. Also, it turns out, that superradiant phases is a general property of the ultrastrong coupling limit~\cite{Felicetti2020}.

In this Letter, we show that the consideration of the atomic mechanical degree of freedom opens the route towards the realization of the ultrastrong coupling regime in the WQED structures. While, the joint dynamics of mechanical and internal degrees of freedom has been considered previously, the analysis relied on the approximations of either classical dynamics of both positions and polarizations of atoms~\cite{chang2013self} or the truncated Hilbert space for the phonons~\cite{manzoni2017designing}. In this Letter, we provide a rigorous mapping from the optomechanical problem to the quantum Rabi model and show that the self-organisation of atoms predicted in the classical picture corresponds to the Rabi-like phase transition known to appear in the ultrastrong coupling regime. Since it has been recently, a tremendous progress in finding analytical solutions of the Rabi model~\cite{Braak2011}, we believe that the presented mapping is of substantial importance for the further developments of the quantum optomechanics in the regime of strong optomechanical coupling.

We consider a system, schematically depicted in Fig.1: $N$ qubits are placed in the laser harmonic traps on top of the chiral ring resonator. 
 \begin{figure}[!h]
     \centering
     \includegraphics[width=1.0\columnwidth]{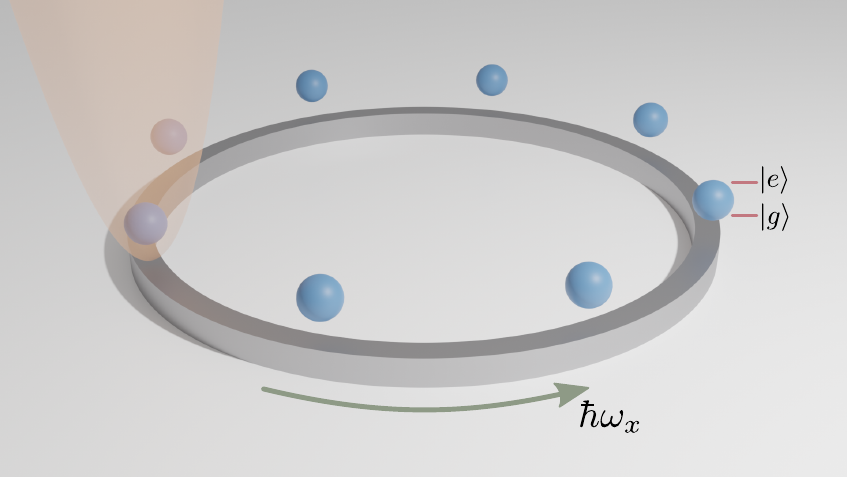}
     \caption{Geometry of the structure: an array of two-level atoms placed in the vicinity of the chiral ring resonator. The parabolic  trapping potential is shown with a shaded region only for one atom.}
     \label{fig:model}
 \end{figure}
The qubit  can absorb or emit a waveguide photon, and the radiative relaxation to the far field is suppressed. The Hamiltonian of the system can be written as
\begin{align}
\hat{H}=\sum_k \omega_k \hat{c}_k^{\dagger}\hat{c}_k +\sum_{j=1}^{N} \omega_x\sigma_j^{+}\sigma_j+\sum_{j=1}^{N}\Omega \hat{a}_j^{\dagger}\hat{a}_j +\hat{H}_{\rm int},
\end{align}
where $\omega_k=v k$ is the dispersion of the chiral waveguide modes, which assumed to be linear, $v$ is the speed of light in the waveguide, $\omega_x$ is the qubit resonance frequency, and $\Omega$ is the optical trap phonon energy, $\hat{a}_j, \hat{a}_j^\dagger$ are annihilation and creation phonon operators respectively. The interaction Hamiltonian is given by
\begin{align}
\hat{H}_{\rm int}=g\sum_{k,j}\left[\sigma_j^{\dagger}\hat{c}_k e^{\mathrm{i} k[R\phi_j+x_j]}+\mathrm{H.c.}\right],
\end{align}
where $g$ is the Rabi splitting, $R$ is the radius of the ring, $x_j$ corresponds to the deviation of the $j$-th atom from its equilibrium position which is equal to $u_0(\hat{a}_j+\hat{a}_j^{\dagger})$, $u_0=\sqrt{\hbar/(2M\Omega)}$ is the quantum of the mechanical motion, where $M$ is the mass of the qubit. 
It should be noted, that  the optical spectrum of the ring is discrete, rather than continuous with the frequency difference between the modes given by $\delta\omega=v/R$. However, for a large resonator, when $v/R\ll \omega_x$ the limit of the continuous spectrum can be employed. 

We then integrate out the waveguide degrees of freedom by performing the Schrieffer-Wolff transform~\cite{SW_transform}, to obtain the effective Hamiltonian up to the second order of the qubit-photon coupling $g$:
\begin{align}
&\hat{H}_{\rm eff}=\sum_j \omega_x\sigma_j^{+}\sigma_j+\sum_j\Omega \hat{a}_j^{\dagger}\hat{a}_j\nonumber\\-&\frac{\Gamma_0}{2}\sum_{ i<j}\left[\mathrm{i}\sigma_i^{+}\sigma_je^{\mathrm{i}qR\phi_{ij}}e^{\mathrm{i}\eta(\hat{a}_i+\hat{a}_i^{\dagger}-\hat{a}_j-\hat{a}_j^{\dagger})}+\mathrm{H.c.}\right],  \label{H_eff}
\end{align}
where $q=\omega_x/v$, $\Gamma_0=g^2/v$ is the radiative decay rate of a single qubit, and $\eta=qu_0$ is the dimensionless optomechanical interaction. In deriving Eq.~\eqref{H_eff} we used the Markov approximation neglecting the frequency dispersion in the phase factor ($k\approx q$). The Markov approximation holds for $R\Gamma_0/v\ll 1$. We note that in stark contrast to the WQED case the resulting Hamiltonian is Hermitian. This is both due to the fact, that unlike the case of an infinite waveguide, our system is a closed one, and because the radiation to the far field has been neglected. The latter approximation can be adopted when the radiative coupling to the waveguide mode $\Gamma_0$ is much stronger than that to the far-field continuum $\Gamma'$. This can be achieved in the photonic crystal waveguide geometries, where $\Gamma_0/\Gamma'>9$ has been experimentally reported~\cite{burgers2019clocked}. 


The qubit excitation energy $\omega_x$ is the largest energy scale of the problem. Since the Hamiltonian commutes with the excitation number operator, We can safely project the Hamiltonian to the subspace with a single excitation. In this case the qubit subspace is spanned by $N$  states, corresponding to excitation localized at each of $N$ qubits. We assume the equidistant spacing of the harmonic traps, i.e. $\phi_{i+1,i}=\phi$. 

The third term in Eq.~\eqref{H_eff} contains the exponent of the bosonic operators making it highly nonlinear in the region $\eta\approx 1$. It is instructive to estimate the experimentally relevant range of parameter values of the model. Parameter $\eta$ is defined by the ratio of the length scale of the mechanical atomic movement, $u_0$ and the wavelength of the photon in the waveguide, $\lambda$, $\eta=4\pi u_0/\lambda$. Parameter $u_0$ can be roughly estimated via the de Broglie wavelength $u_0<\hbar/p_{\rm th}$, where the thermal momentum $p_{\rm th}=\sqrt{3Mk_BT}$. For the lithium atoms and the resonant wavelength approximately 700 nm the value of $\eta=1$ is achieved at $T=640$ nK, which is a  temperature which has been achieved in recent cold atom experiments (see the review~\cite{anglin2002bose} and references within). The corresponding phonon energy is then approximately $2.4$ kHz. The radiative decay rate $\Gamma_0$ can be flexibly tuned in a wide range of frequencies from zero to the GHz. Therefore, the range of $\Gamma_0/\Omega,\eta \sim 1$ can be achieved in the state of the art cold atom experiments. Thus, it is relevant to explore the properties of Hamiltonian~\eqref{H_eff} outside the small $\eta$ regime.

We introduce the unitary transformation $T_N$ for the case of $N$ qubits which transforms Eq.~\eqref{H_eff} to a more familiar form. The general expression for $T_N$ can be found in SI. For the case of two qubits, $T_2$ reads
$\hat{T}\hat{H}_{eff}\hat{T}^{\dagger}$, where
\begin{align}
    T=\frac{1}{\sqrt{2}}\begin{pmatrix} \mathrm{i}e^{-\mathrm{i}\eta \hat{x}_1} & e^{-\mathrm{i}\eta \hat{x}_2 - \mathrm{i}qR\phi} \\ -\mathrm{i}e^{\mathrm{i}\eta \hat{x}_1} & e^{-\mathrm{i}\eta \hat{x}_2 - \mathrm{i} qR\phi}\end{pmatrix},
\end{align}
where $\hat{x}_i=\hat{a}_i+\hat{a}_i^{\dagger}$, and the transformed Hamiltonian 
\begin{align}
    &\hat{T_2}\hat{H}_{eff}\hat{T_2}^{\dagger}=\Omega \left[ \hat{a}_{CM}^{\dagger}\hat{a}_{CM}+ \hat{a}_{d}^{\dagger}\hat{a}_{d}+\frac{\eta^2}{2}\right.\nonumber\\&\left.+\sigma_x \frac{\eta}{\sqrt{2}}(\mathrm{i}\hat{a}_d - \mathrm{i}\hat{a}_d^{\dagger})-\frac{\Gamma_0}{2\Omega}\sigma_z\right],
\end{align}
$\hat{a}_{CM}=\frac{1}{\sqrt{2}}(\hat{a}_1+\hat{a}_2+i\eta)$ corresponds to the centre of mass qubit motion and $\hat{a}_d= \frac{1}{\sqrt{2}}(\hat{a}_1-\hat{a}_2)$ corresponds to the relative motion of two qubits. We first note, that the centre of mass motion operator is shifted from the equilibrium position on $\eta$. This is due to the unidirectional propagation of the chiral waveguide photon, which \textit{pushes} the qubits as whole in one direction. Then, we see that the spectrum of the problem does not depend on the static phase difference $\phi$, which is typical for the chiral waveguide quantum optical set-ups~\cite{pletyukhov2012scattering, kornovan2017transport}. Finally, we see that up-to the centre-of-mass kinetic energy term, which decouples from the rest of the system, the effective Hamiltonian is exactly the one corresponding to the quantum Rabi model.  The radiative decay $\Gamma_0$ plays the role of the resonant transition energy and the dimensionless optomechanical coupling defines the effective coupling strength. The case of strong optomechanical interaction $\eta>0.1\sqrt{2}$  thus directly maps to the ultrastrong coupling regime (USC). It is known that in the USC and deep-strong coupling regime ($\eta>\sqrt{2}$) of the Rabi model, the system is characterized by the non-vacuum ground state $|\Psi_G\rangle$ which can be roughly approximated by the superposition of the coherent states $|\Psi_G\rangle\approx \frac{1}{\sqrt{2}}(|+\rangle\otimes |\alpha\rangle+|-\rangle\otimes |-\alpha\rangle)$~\cite{emary2004phase}, where  $|\pm\alpha\rangle$ are the bosonic coherent states, and $|\pm\rangle =\frac{1}{\sqrt{2}}(|\uparrow\rangle\pm|\downarrow\rangle)$ - are the superpositions of the ground and excited qubit states. We note also that the direct mapping to the Rabi model is valid only in the purely chiral case. However, as we show in Supplemental material, the numerically obtained spectrum for the non-perfectly chiral waveguide qualitatively is very similar to the perfectly chiral case.

For three qubits, the unitary transformation $T_3$ results in the Hamiltonian (see details in Supplemental material~\cite{supp_mat}):
\begin{align}
    &\hat{T_3}\hat{H}_{eff}\hat{T_3}^{\dagger}=\hat{\tilde{H}}_{eff}=\hat{H}_{ph}+\hat{H}_q+\hat{H}_c, \label{fullH3}
\end{align}
where $\hat{H}_{ph}$ is the phonon kinetic energy given by
\begin{align}
  \hat{H}_{ph}=\Omega(\hat{a}^{\dagger}\hat{a}+\hat{a}_x^{\dagger}\hat{a}_x+\hat{a}_y^{\dagger}\hat{a}_y+\frac{2\eta^2}{3}),
\end{align}
where $\hat{a}$ corresponds to the shifted operator of centre-of-mass motion, $\hat{a}=\frac{1}{\sqrt{3}}(\hat{a}_1+\hat{a}_2+\hat{a}_3+\mathrm{i}\eta)$, and $\hat{a}_x= \frac{1}{\sqrt{6}}(-\hat{a}_1-\hat{a}_2+2\hat{a}_3)$,  $\hat{a}_{y}=\frac{1}{\sqrt{2}}(\hat{a}_1-\hat{a}_2)$ are operators of normal modes. The qubit Hamiltonian $\hat{H}_q$ reads
\begin{align}
    \hat{H}_q=-\frac{\sqrt{3}\Gamma_0}{2}\hat{\lambda}_3,
\end{align}
where $\hat{\lambda}_i$ is the $3\times3$ Gell-Mann matrix. Finally, the coupling term $\hat{H}_c$ is given by 
\begin{align}
\hat{H_c}=-\frac{\Omega\eta}{\sqrt{3}}\left[\hat{p}_x(\hat{\lambda}_1+\hat{\lambda}_4+\hat{\lambda}_6)+\hat{p}_y(-\hat{\lambda}_2+\hat{\lambda}_5-\hat{\lambda}_7)\right],
\end{align}
where $\hat{p}_i=\frac{\mathrm{i}}{\sqrt{2}}(\hat{a}_i-\hat{a}_i^{\dagger})$. We note that, the Hamiltonian $\hat{\tilde{H}}_{eff}$ (up to the decoupled centre of mass motion) describes the two-dimensional Bose condensate (BEC) of spin 1 particles, localized in a harmonic trap (given by $\hat{H}_{ph}$) and in perpendicular magnetic field ($\hat{H}$_q). The term $\hat{H}_c$ describes the spin-orbit coupling (SOC) for spin 1 particles. This type SOC has been introduced for the BECs of spin particles previously~\cite{barnett20123,PhysRevA.94.033629}. Thus, here we highlight a link between the waveguide optomechanical systems and BEC physics.

We first note, that despite seeming similarity, the Hamiltonian in Eq.~\eqref{fullH3} is qualitatively different from the Dicke model Hamiltonian. Namely, the qubit operators do not obey the angular momentum commutation relations. Moreover, the Hamiltonian~\eqref{fullH3} possesses global $\mathbb{Z}_3$ symmetry. Consider the unitary operator
\begin{align}
    \hat{R}=e^{-\mathrm{i}\hat{L}_z \frac{2\pi}{3}}\otimes\begin{pmatrix} 1 & 0 & 0 \\ 0 & e^{\mathrm{i}4\pi/3} & 0 \\ 0 & 0 & e^{\mathrm{i}2\pi/3}\end{pmatrix},
\end{align}
where $\hat{L}_z=\hat{x}\hat{p}_y-\hat{y}\hat{p}_x$ is the angular momentum operator. Operator $\hat{R}$ obeys $\hat{R}^2=\hat{R}^{\dagger}$ and thus $\left[\mathds{1},\hat{R},\hat{R}^2\right]$ form a group. We note that $\hat{R}\hat{\tilde{H}}_{eff}\hat{R}^{\dagger}=\hat{\tilde{H}}_{eff}$ and thus $[\hat{R},\hat{\tilde{H}}_{eff}]=0$. Therefore, the eigenstates of $\hat{R}$ are also eigenstates of $\hat{\tilde{H}}_{eff}$. The three distinct eigenvalues of $\hat{R}$ are $\left[1,e^{\mathrm{i}2\pi/3},e^{\mathrm{i}4\pi/3}\right]$.

We then assume the limit of the classical motion of the qubits by assuming $\hat{p}_x,\hat{p}_y$ to be classical variables and find the eigenvalues of the corresponding matrix Hamiltonian obtained from Eq.~\eqref{fullH3}. We find the ground state energy by minimizing the smallest eigenvalue with respect to $p_x,p_y$. Moving to the polar coordinates $(p_x,p_y)=(p\cos\theta,p\sin\theta)$ we find that the minimum energy is obtained for $\cos 3\theta=1$. With this condition fulfilled, the expression for the ground state energy as a function of $p$ reads
\begin{align}
    \epsilon_G=\frac{2\eta^2\Omega}{3}+\frac{\sqrt{3}\Gamma_0}{2}\left[\frac{\tilde{p}^2}{2\mu}-2(\tilde{p}^2+\frac{1}{3})^{\frac{1}{2}}\cos\left(\frac{\gamma}{3}\right)\right],\label{ee_G}
\end{align}
where $\mu=\sqrt{4/27}\eta^2\Omega/\Gamma_0$, $\tilde{p}=2\eta\Omega/(3\Gamma_0)p$, and $\gamma=\arctan\left((81\tilde{p}^4+27\tilde{p}^2+3)^{\frac{1}{2}}/9\tilde{p}^3\right)$. For small $\tilde{p}$ we can write
\begin{align}
    \epsilon_G\approx \frac{2\eta^2\Omega}{3}+\frac{\sqrt{3}\Gamma_0}{2}\left[-1-\tilde{p}^3+\frac{9\tilde{p}^4}{8}+\frac{\mu-3}{2\mu}\tilde{p}^2\right] \label{epsilon_G_approx}
\end{align}
For $\eta\ll 1$ Eq.~\eqref{epsilon_G_approx} has a single  local minimum at $\tilde{p}=0$. For $\eta>\eta_c=\sqrt{3\sqrt{3}\Gamma_0/(7\Omega)}$ it has an additional minimum at $\tilde{p}_c$ which for $\eta\approx\eta_c$ can by approximated by  $\tilde{p}_c\approx \frac{1}{3}(1+\sqrt{7-2/\mu})$. Then, for $\eta>\sqrt{\sqrt{3}\Gamma_0/(2\Omega)}$, there is only a single minimum at $\tilde{p}_c$. The situation, when there is a range of the parameters where both phases coexist is characteristic for the first order  quantum phase transitions. Indeed, at $\eta=\eta_c$ the first derivative of $\epsilon_G$ is discontinous which is a hallmark of the first order quantum phase transition~\cite{sachdev2007quantum}.

\begin{figure}[!h]
     \centering
     \includegraphics[width=1.0\columnwidth]{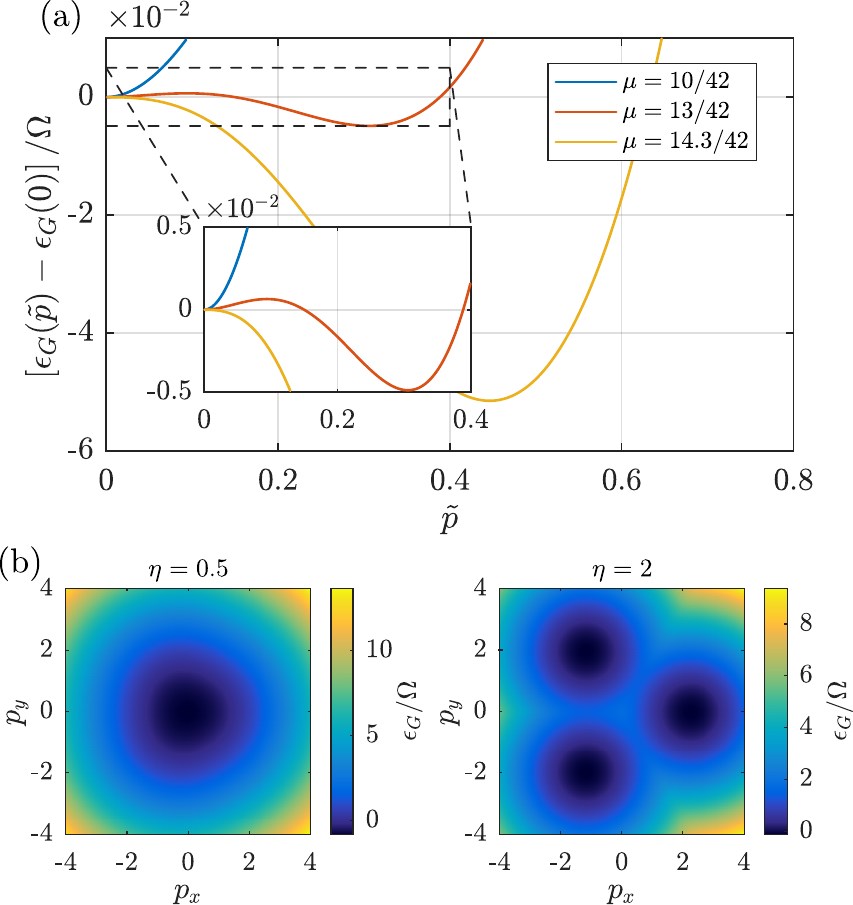}
     \caption{(a) Dependency of the ground state energy on $\tilde{p}$ for different values of the parameter $\mu$; $\Gamma_0 / \Omega = 2.5$. (b) Dispersion of the lowest energy surface in the classical approximation for qubit motion in the two cases: $\eta=0.5$ and $\eta=2.0$; $\Gamma_0/\Omega=2.5$, $\eta_c \approx 1.36$.}
     \label{fig:Eground}
 \end{figure}

We plot dependence of $\epsilon_G$ given by Eq.~\eqref{ee_G} in Fig.~\ref{fig:Eground}(a). We can see, that indeed there exists a range of parameters where there are two local minima signifying the phase co-existence regime. Thus, the quantum phase transition (QPT) in the classical limit is indeed of the first order.  This is in stark contrast to the classical limit of quantum Rabi model, where the phase transition is of the second order~\cite{PhysRevLett.115.180404}.

The map of the $\epsilon_G$ in $(p_x,p_y)$ space is shown in Fig.~\ref{fig:Eground}(b). For the case $\eta<\eta_c$ shown in left panel,  there is a single minimum corresponding to $\tilde{p}=0$. For $\eta>\eta_c$ (right) panel three degenerate minima emerge.

 Since the QPTs can occur only in the thermodynamic limit we shall refine our analysis of the ground state energy. For that we first consider that the actual quantum states corresponding to the minimal energy in the classical limit are the direct products of the spin states and the coherent states of the qubit motion at small $p_c$
 \begin{align}
|l\rangle \approx\mathcal{N}_c \begin{pmatrix}\tilde{p}_c/2 \\ -[1-\frac{5}{8}\tilde{p}_c^2]e^{2i\theta_l} \\ \tilde{p}_c e^{i\theta_l}\end{pmatrix}\otimes \left|\tilde{p}_c\cos \theta_l,\tilde{p}_c\sin \theta_l\right.\left.\right>,
\end{align}
where $l=0,1,2$, $\theta_l=2\pi l/3$, and  $\mathcal{N}_c$ is the normalization factor. It is evident that $\langle l | \hat{\tilde{H}}_{eff}|l\rangle$ yields the classical {mean-field} ground state energy. However, these states can not be the eigenstates of Hamiltonian $\hat{\tilde{H}}_{eff}$ since they are not eigenstates of operator $\hat{R}$. Namely, $\hat{R}|l\rangle=|[(l+1) \mathrm{mod} 3]\rangle$. Moreover, due to the nonorthogonality of the coherent states $\langle l' |\hat{\tilde{H}}_{eff}|l\rangle \neq E\delta_{l',l}$ and $\langle l' |l\rangle \neq \delta_{l',l}$. We thus can solve the characteristic equation for the eigenvalues 
$\mathrm{det}[\langle l' |\hat{\tilde{H}}_{eff}|l\rangle-E\langle l' |l\rangle]=0$. The explicit form of the characteristic equation is cumbersome and presented in Supplemental material~\cite{supp_mat}. It is however important to note, that the non-diagonal elements of the matrix representation of the Hamiltonian are proportional to the overlap of the coherent states which is proportional to $\mathrm{exp}[-3\tilde{p}_c^2]$, and thus the splitting decreases rapidly as we depart from the phase transition at $\eta_c$. The explicit form of the eigenstates can be found from the symmetry considerations. Namely, the eigenstates should also be the eigenstates of the operator $\hat{R}$. We then can easily find the mutually orthogonal linear superpositions of states $|l\rangle$ which satisfy this condition. Namely, the ground and two excited states are given by:
 \begin{align}
    & |\Psi_G\rangle = \frac{1}{\sqrt{3}}\left[|0\rangle + |1\rangle +|2\rangle\right],\nonumber\\
    & |\Psi_{E1}\rangle = \frac{1}{\sqrt{3}}\left[|0\rangle + e^{4\mathrm{i}\pi/3}|1\rangle +e^{2\mathrm{i}\pi/3}|2\rangle\right],\nonumber\\
    & |\Psi_{E2}\rangle = \frac{1}{\sqrt{3}}\left[|0\rangle + e^{2\mathrm{i}\pi/3}|1\rangle +e^{4\mathrm{i}\pi/3}|2\rangle\right]. \label{eq:Estates}
 \end{align}
 
\begin{figure}[!t]
    \centering
    \includegraphics{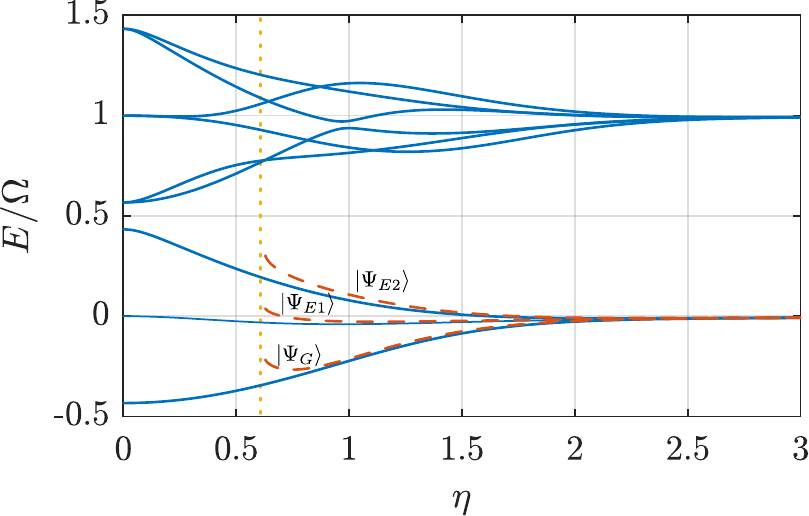}
    \caption{Eigenergies of first nine eigenstates of $\hat{\tilde{H}}_{eff}$ vs optomechanical coupling $\eta$ with $\Gamma_0 / \Omega = 0.5$. Dashed lines show the dispersions of states in Eq.~\eqref{eq:Estates}. Vertical dotted line corresponds to critical optomechanical coupling $\eta_c \approx 0.61$. For the numerical diagonalization, the phonon sub-space was truncated with maximal phonon occupation number - 100.} 
    \label{fig:spectrum}
\end{figure}

The spectrum of $\hat{\tilde{H}}_{eff}$ as a function of the coupling strength $\eta$ is shown in Fig.~\ref{fig:spectrum} for the case of the ground state of the centre of mass degree of freedom $\hat{n}_{CM}=0$. The spectrum has been obtained via the direct numerical diagonalization by truncating the phonon subspace. We can see that at large $\eta$ the ground state becomes quasi-degenerate. We also plot the analytically obtained dispersions of states $|\Psi_G\rangle,|\Psi_{E1}\rangle,|\Psi_{E2}\rangle$. 
As can be seen, the first three low energy states  given by Eq.~\eqref{eq:Estates} are the analog of the triangular  Schrodinger cat states~\cite{vlastakis2013deterministically}. While the Schrodinger cat states are generally regarded as extremely fragile with respect to decoherence, it has been recently revealed that the two-component cat states appearing in the USC of the conventional Rabi model appear two be robust to decoherence and can be used to realize protected quantum gates with high fidelity~\cite{PhysRevLett.107.190402,wang2016holonomic}. Thus, the states $|\Psi_{[G,E1,E2]}\rangle$ as the three-component generalizations of the cat states originating in the USC are likely to remain sufficiently stable and can be used for quantum information processing.

We have shown, that the phase transition occurs in the classical limit. As has been shown recently for the quantum Rabi problem, the classical limit can be regarded as a thermodynamic limit of the vanishing harmonic oscillator energy $\Omega$~\cite{PhysRevA.85.043821,PhysRevA.87.013826,PhysRevLett.115.180404,puebla2017probing}. To explore this limit in our case, we redefine the energy constants in $\hat{\tilde{H}}_{eff}$ in the following way: we set $\eta\Omega\rightarrow \eta'$ as an independent variable and redefine $\Gamma_0=\xi\omega$, $\Omega=\omega/\xi$. The thermodynamic limit is then achieved for $\xi\rightarrow\infty$. 
\begin{figure}[!h]
    \centering
    \includegraphics[width=1.0\columnwidth]{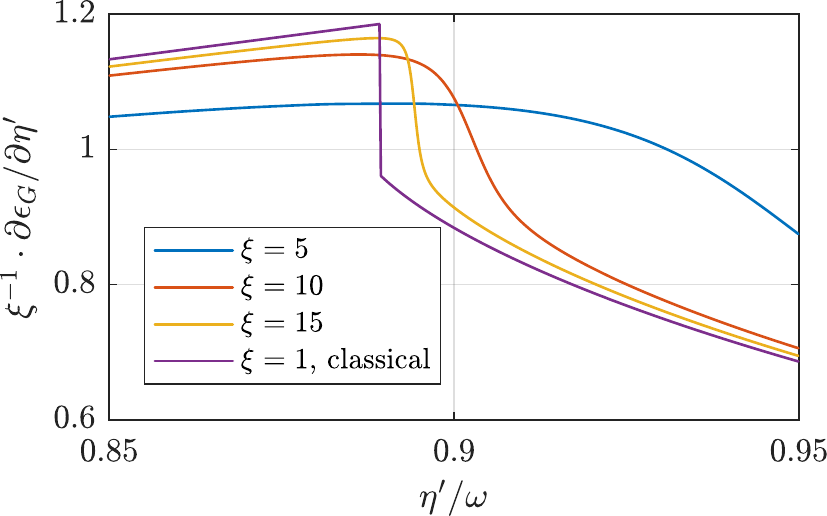}
    \caption{First derivative of the ground state energy $\partial\epsilon_G/\partial \eta'$ for different values of scaling parameter $\xi$. $\omega=1$.}  
    \label{fig:QPT}
\end{figure}

In Fig.~\ref{fig:QPT} we plot the first derivative of the ground state energy as a function of $\eta'$ for $\omega=1$ and for different $\xi$. We can see that as $\xi$ increases this function steepens in the vicinity of $\eta_c'$. In the limit of infinite $\xi$ we would observe the discontinuity of the $\partial\epsilon/\partial \eta'$ just as in the classical limit and the establishment of the QPT with $\mathbb{Z}_3$ symmetry breaking. The $\Omega\rightarrow 0$ limit can be regarded as the classical limit of the atomic motion. Thus, the predicted phase transition corresponds to the displacement of atoms with respect to the centres of the traps, i.e. the self-organisation of atoms due to the photon mediated inter-atomic interactions. The self-organization of atoms has been predicted within the classical approach in WQED systems~\cite{chang2013self}. We thus reveal the direct connection of the self organization phenomena and quantum phase transitions similar to that occurring in the Rabi model.
       
To conclude, we have established a direct mapping between the quantum optomechanical set-up in the chiral  waveguide and the generalization of the quantum Rabi-model. While for two qubits, the system directly maps to the quantum Rabi model, already for the case of three qubits, the system possesses unconventional $\mathbb{Z}_3$  symmetry, exhibiting multi-component Schrodinger-cat ground states as well as $\mathbb{Z}_3$ symmetry breaking first order  phase transitions in the thermodynamic limit.  The work establishes solid connections between the self-organization of atoms in photonic structures which has been previously treated and quantum phase transitions. It also poses an interesting question on the structure of the ground state in the limit of the large number of qubits $N$. While we have demonstrated the $\mathbb{Z}_N$ symmetry for $N$ qubits (see SI), the nature of the phase transition and the structure of the ground state is yet to be explored.   

The results of the paper can be applied to a more general class of systems of moving atoms in the photonic structures, since it reveals that the apparatus developed in the studies of USC can be directly applied to explore both fundamental aspects of quantized spin-motion coupling and perspective applications in quantum information processing. 

\section{acknowledgements}
The work of Ivan Iorsh (mapping to the Rabi model) was supported by the Russian Science Foundation (project 20-12-00224). The work of Denis Sedov (numerical solution of three qubit model) was supported by Russian Science Foundation (project 20-12-00194).  We thank A.N. Poddubny, A.V. Poshakinskiy, and M.I. Petrov for fruitful discussions.
\bibliography{OptMech_Quant}

\newpage
\onecolumngrid
\setcounter{page}{0}
\setcounter{table}{0}
\setcounter{section}{0}
\setcounter{figure}{0}
\setcounter{equation}{0}
\renewcommand{\thepage}{\Roman{page}}
\renewcommand{\thesection}{S\arabic{section}}
\renewcommand{\thetable}{S\arabic{table}}
\renewcommand{\thefigure}{S\arabic{figure}}
\renewcommand{\theequation}{S\arabic{equation}}
\cleardoublepage
\vfill\eject
\thispagestyle{empty}
\newcommand{\comm}[2]{\left[ #1,\, #2  \right]}
\newcommand*\mycommand[1]{\texttt{\emph{#1}}}
\newcommand{\micron}{\hbox{\textmu}\text{m}}
\newcommand{\figurewidth}{0.85\columnwidth}
\graphicspath{ {./figures/} }

\section*{Supplemental Material}

\subsection{Mapping to the $N$-state quantum Rabi model}
We started from the Hamiltonian in Eq.~(3) from the main text projected on the subspace of a single qubit excitation
\begin{align}
    \hat{H}_{eff}=\hat{H}_{phon}+\hat{H}_{c}=\sum_{j=1}^{N} \Omega \hat{a}_j^{\dagger}\hat{a}_j-\frac{\Gamma_0}{2}\sum_{ (i<j)}^{N} \left[\mathrm{i} |i\rangle \langle j| e^{\mathrm{i} qR\phi(i-j)}e^{\mathrm{i} qu_0(\hat{a}_i+\hat{a}_i^{\dagger}-\hat{a}_j-\hat{a}_j^{\dagger})}+\mathrm{H.c.}\right],
\end{align}
where $|i\rangle$ are the orth vectors in $N-$ dimensional space corresponding to the excitation localized at the $i$-th qubit.
We then introduce the unitary operator $\hat{T}= \hat{S}\times\hat{\Phi}$, where 
\begin{align}
    \hat{\Phi}_{ij}=\delta_{ij} e^{-\mathrm{i}qR\phi(i-1)}e^{-\mathrm{i}qu_0(\hat{a}_{i}+\hat{a}_{i}^{\dagger})},\quad i,j=1\ldots N,
\end{align}
and $\hat{S}$ is the matrix, which lines are the normalized eigenvectors of the matrix $\sum_{j>i}\left[-\mathrm{i}|j\rangle\langle i |\right] +H.c.$
The unitary transformation $\hat{T}$ diagonalizes $\hat{H}_c$. The resulting diagonal matrix has non-degenerate eigenvalues symmetric with respect to zero. For $N=2$, eigenvalues are $\pm \Gamma_0/2$.

The transformation $\hat{\Phi} \hat{H}_{phon} \hat{\Phi}^{\dagger}$ results in 
\begin{align}
\hat{\Phi} \hat{H}_{phon} \hat{\Phi}^{\dagger}=\sum_{j=1}^{N} \Omega \hat{a}_j^{\dagger}\hat{a}_j +\Omega\eta^2 +\Omega\eta\times \mathrm{diag}[\mathrm{i}(\hat{a}_i^{\dagger}-\hat{a}_i)], \label{eq:H_after_Phi}
\end{align}
where $\eta=qu_0$. We then can introduce the phonon
\textit{centre of mass}  creation operator $\hat{a}_{CM}=\frac{1}{\sqrt{N}}\sum_{i}\hat{a}_i$. Namely, for $N=2$ the resulting Hamiltonian reads:
\begin{align}
    \Omega \hat{a}_{CM}^{\dagger}\hat{a}_{CM}+\Omega a_{d}^{\dagger}a_{d}-\Omega  i\frac{\eta}{\sqrt{2}} (\hat{a}_{CM}-\hat{a}_{CM}^{\dagger})+\Omega\eta^2+\sigma_x \Omega\frac{\eta}{\sqrt{2}}(i\hat{a}_d-i\hat{a}_d^{\dagger})-\frac{\Gamma_0}{2}\sigma_z,
\end{align}
where $\hat{a}_d=(\hat{a}_1-\hat{a}_2)/\sqrt{2}$ corresponds to the relative motion of the two qubits. As can be seen this Hamiltonian is exactly the one describing the Quantum Rabi model plus decoupled bosonic mode corresponding to the centre of mass motion. Now we introduce new variables 
\begin{align}
	\hat{a} = \hat{a}_{CM} + \mathrm{i}\frac{\eta}{\sqrt{2}},\quad \hat{a}^\dagger = \hat{a}_{CM}^\dagger - \mathrm{i} \frac{\eta}{\sqrt{2}},
\end{align}
for which we have 
\begin{equation}
    \Omega \hat{a}_{CM}^{\dagger}\hat{a}_{CM}-\Omega  i\frac{\eta}{\sqrt{2}} (\hat{a}_{CM}-\hat{a}_{CM}^{\dagger})=\Omega \hat{a}^{\dagger}\hat{a}-\frac{\Omega\eta^2}{2}
\end{equation}
thus finally we arrive at
\begin{align}
    \hat{\tilde{H}}_{\text{eff}}=T\hat{H}_{eff} T^{\dagger}=\Omega \hat{a}^{\dagger}\hat{a}+\Omega \hat{a}_{d}^{\dagger}\hat{a}_{d}+\frac{\Omega\eta^2}{2}+\sigma_x \Omega\frac{\eta}{\sqrt{2}}(i\hat{a}_d-i\hat{a}_d^{\dagger})-\frac{\Gamma_0}{2}\sigma_z,
\end{align}

For $N=3$ qubits the Hamiltonian is given by
\begin{align}
    &\hat{\tilde{H}}_{\text{eff}}=T\hat{H}_{eff} T^{\dagger}=\Omega \hat{a}_{CM}^{\dagger}\hat{a}_{CM}-\Omega  i\frac{\eta}{\sqrt{3}} (\hat{a}_{CM}-\hat{a}_{CM}^{\dagger})+\Omega (\hat{a}_{x}^{\dagger}\hat{a}_{x}+\hat{a}_{y}^{\dagger}\hat{a}_{y})+\Omega\eta^2-\nonumber\\&\frac{\sqrt{3}\Gamma_0}{2}\begin{pmatrix}1 & 0 & 0 \\ 0 & -1 & 0 \\ 0 & 0 & 0  \end{pmatrix} -     \Omega\frac{\eta}{\sqrt{6}}(i\hat{a}_{x}-i\hat{a}_{x}^{\dagger})\begin{pmatrix} 0 & 1 & 1 \\ 1 & 0 & 1 \\ 1  & 1 & 0  \end{pmatrix} +\Omega\frac{\eta}{\sqrt{6}}(\hat{a}_{y}-\hat{a}_{y}^{\dagger})\begin{pmatrix} 0 & 1 & -1 \\ -1 & 0 & 1 \\ 1  & -1 & 0  \end{pmatrix},
\end{align}
where $\hat{a}_{x}= \frac{1}{\sqrt{6}}(-\hat{a}_1-\hat{a}_2+2a_3), \quad \hat{a}_{y}=\frac{1}{\sqrt{2}}(\hat{a}_1-\hat{a}_2)$.
Now we introduce new variables 
\begin{align}
	\hat{a} = \hat{a}_{CM} + \mathrm{i}\frac{\eta}{\sqrt{3}},\quad \hat{a}^\dagger = \hat{a}_{CM}^\dagger - \mathrm{i} \frac{\eta}{\sqrt{3}},
\end{align}
for which we have 
\begin{equation}
    \Omega \hat{a}_{CM}^{\dagger}\hat{a}_{CM}-\Omega  i\frac{\eta}{\sqrt{3}}   (\hat{a}_{CM}-\hat{a}_{CM}^{\dagger})=\Omega \hat{a}^{\dagger}\hat{a}-\frac{\Omega\eta^2}{3}
\end{equation}
thus finally we arrive at
\begin{align}
    &\hat{\tilde{H}}_{\text{eff}}=T\hat{H}_{eff} T^{\dagger}=\Omega \hat{a}^{\dagger}\hat{a}+\Omega (\hat{a}_{x}^{\dagger}\hat{a}_{x}+\hat{a}_{y}^{\dagger}\hat{a}_{y})+\Omega\frac{2\eta^2}{3}-\nonumber\\&\frac{\sqrt{3}\Gamma_0}{2}\begin{pmatrix}1 & 0 & 0 \\ 0 & -1 & 0 \\ 0 & 0 & 0  \end{pmatrix} -     \Omega\frac{\eta}{\sqrt{6}}(i\hat{a}_{x}-i\hat{a}_{x}^{\dagger})\begin{pmatrix} 0 & 1 & 1 \\ 1 & 0 & 1 \\ 1  & 1 & 0  \end{pmatrix} +\Omega\frac{\eta}{\sqrt{6}}(\hat{a}_{y}-\hat{a}_{y}^{\dagger})\begin{pmatrix} 0 & 1 & -1 \\ -1 & 0 & 1 \\ 1  & -1 & 0  \end{pmatrix}.
\end{align}
In general we a generalized Rabi model with $N$-dimensional matrices.

\subsection{Hamiltonian, projected on the three lowest energy states}
In this subsection we calculate the 3 by 3 matrix ($l,l'=0,1,2$) $\langle l'\lvert\hat{\tilde{H}}_{\text{eff}}\rvert l\rangle$ -- i.e. the projection of our Hamiltonian $\hat{\tilde{H}}_{\text{eff}}$ on the three lowest energy states. The states $\rvert l\rangle$ are given by Eq.~13 from the main text, but it is important to note that in this equation one has to interpret the phonon state $\lvert\tilde{p}_c\cos \theta_l,\tilde{p}_c\sin \theta_l\rangle$ (with $\theta_l=2\pi l/3$) as a product of coherent states of x- and y-phonons, namely, $\lvert\tilde{p}_c\cos \theta_l,\tilde{p}_c\sin \theta_l\rangle=\lvert\alpha^{(l)}_{c;x}\rangle\otimes\lvert\alpha^{(l)}_{c;y}\rangle$. The coherent states satisfy $\hat{a}_{x}\lvert\alpha^{(l)}_{c;x}\rangle=\alpha^{(l)}_{c;x}\lvert\alpha^{(l)}_{c;x}\rangle$ and $\hat{a}_{y}\lvert\alpha^{(l)}_{c;y}\rangle=\alpha^{(l)}_{c;y}\lvert\alpha^{(l)}_{c;y}\rangle$. How are the parameters  $\alpha^{(l)}_{c;x}$ and $\alpha^{(l)}_{c;y}$ of the coherent states related to the pair of numbers $(\tilde{p}_c\cos \theta_l,\tilde{p}_c\sin \theta_l)$? We remind, that $\hat{p}_i=\frac{\mathrm{i}}{\sqrt{2}}(\hat{a}_i-\hat{a}_i^{\dagger})$ and $\hat{x}_i=\frac{1}{\sqrt{2}}(\hat{a}_i+\hat{a}_i^{\dagger})$, here $i=x,y$. Now, by definition $\langle \alpha^{(l)}_{c;x}\lvert \hat{p}_x\rvert\alpha^{(l)}_{c;x}\rangle=p_x=p_c\cos{\theta_l}$, $\langle \alpha^{(l)}_{c;y}\lvert \hat{p}_y\rvert\alpha^{(l)}_{c;y}\rangle=p_y=p_c\sin{\theta_l}$ and we remind that $\tilde{p}_c=2\eta\Omega/(3\Gamma_0)p_c$. On the other hand, $p_x=\langle \alpha^{(l)}_{c;x}\lvert \hat{p}_x\rvert\alpha^{(l)}_{c;x}\rangle=\frac{\mathrm{i}}{\sqrt{2}}(\alpha^{(l)}_{c;x}-\alpha^{(l)*}_{c;x})$ and $p_y=\langle \alpha^{(l)}_{c;y}\lvert \hat{p}_y\rvert\alpha^{(l)}_{c;y}\rangle=\frac{\mathrm{i}}{\sqrt{2}}(\alpha^{(l)}_{c;y}-\alpha^{(l)*}_{c;y})$. Also, we can introduce $x_c$ and $y_c$ as $x_c=\langle \alpha^{(l)}_{c;x}\lvert \hat{x}\rvert\alpha^{(l)}_{c;x}\rangle=\frac{1}{\sqrt{2}}(\alpha^{(l)}_{c;x}+\alpha^{(l)*}_{c;x})$ and $y_c=\langle \alpha^{(l)}_{c;y}\lvert \hat{y}\rvert\alpha^{(l)}_{c;y}\rangle=\frac{1}{\sqrt{2}}(\alpha^{(l)}_{c;y}+\alpha^{(l)*}_{c;y})$, and thus we now have a simple linear one-to-one map $\alpha^{(l)}_{c;x}=(-\mathrm{i}p_c\cos{\theta_l}+x_c)/\sqrt{2}$ and $\alpha^{(l)}_{c;y}=(-\mathrm{i}p_c\sin{\theta_l}+y_c)/\sqrt{2}$, which, in other words, means that to fully characterize a coherent state we need to know the average position and the average momentum in this coherent state. However, in the states $\lvert l\rangle$ the coherent parameters $\alpha^{(l)}_{c;x}$ and $\alpha^{(l)}_{c;y}$ are pure imaginary, and thus $x_c=y_c=0$.  We now proceed with calculating the matrix elements, assuming that  $\lvert l\rangle=A^{(l)}_c\otimes\lvert\alpha^{(l)}_{c;x}\rangle\otimes\lvert\alpha^{(l)}_{c;y}\rangle=A^{(l)}_c\otimes\lvert -\mathrm{i}p_c\cos{\theta_l}/\sqrt{2}\rangle\otimes\lvert-\mathrm{i}p_c\sin{\theta_l}/\sqrt{2}\rangle$, where $A^{(l)}_c$ is the corresponding column-vector in Eq.~13 (from the main text) with the normalization factor $\mathcal{N}_c$. In our derivation below we use that the overlap of coherent states is given by $\langle\beta\rvert\alpha\rangle=e^{-(|\beta|^2+|\alpha|^2-2\beta^{*}\alpha)/2}$. And we use the fact, that our Hamiltonian is written in the normal-ordered form $\hat{\tilde{H}}_{\text{eff}}[\hat{a}_x,\hat{a}_x^{\dagger},\hat{a}_y,\hat{a}_y^{\dagger}]=:\hat{\tilde{H}}_{\text{eff}}[\hat{a}_x,\hat{a}_x^{\dagger},\hat{a}_y,\hat{a}_y^{\dagger}]:$, i.e. in each term, each creation operator is to the left of each annihilation operator, which greatly simplifies "sandwiching" the Hamiltonian between coherent states. We arrive at the following result (we omit the center-of-mass term)
\begin{align}
    &\langle l'\lvert\hat{\tilde{H}}_{\text{eff}}\rvert l\rangle=\nonumber\\
    &A^{(l')\dagger}_c\otimes\langle \frac{-\mathrm{i}p_c\cos{\theta_{l'}}}{\sqrt{2}}\lvert\otimes\langle\frac{-\mathrm{i}p_c\sin{\theta_{l'}}}{\sqrt{2}}\lvert \hat{\tilde{H}}_{\text{eff}}A^{(l)}_c\otimes
    \lvert \frac{-\mathrm{i}p_c\cos{\theta_l}}{\sqrt{2}}\rangle\otimes\lvert\frac{-\mathrm{i}p_c\sin{\theta_l}}{\sqrt{2}}\rangle
    =\nonumber\\
    &A^{(l')\dagger}_c\hat{\tilde{H}}_{\text{eff}}[\hat{a}_x\rightarrow\alpha^{(l)}_{c;x},\hat{a}_x^{\dagger}\rightarrow\alpha^{(l')*}_{c;x},\hat{a}_y\rightarrow\alpha^{(l)}_{c;y},\hat{a}_y^{\dagger}\rightarrow\alpha^{(l')*}_{c;y}]A^{(l)}_c=\nonumber\\
    &\begin{pmatrix}
        &\tilde{H}_{\text{eff}, 00} 
        &\tilde{H}_{\text{eff}, 01} 
        &\tilde{H}_{\text{eff}, 01}^{*} \\ 
        &\tilde{H}_{\text{eff}, 01}^{*}  &\tilde{H}_{\text{eff}, 00} 
        &\tilde{H}_{\text{eff}, 01}\\
        &\tilde{H}_{\text{eff}, 01}
        &\tilde{H}_{\text{eff}, 01}^{*}
        &\tilde{H}_{\text{eff}, 00} 
    \end{pmatrix},
\end{align}
and we remind that $\alpha^{(l)}_{c;x}=-\mathrm{i}p_c\cos{\theta_l}/\sqrt{2}$ and $\alpha^{(l)}_{c;y}=-\mathrm{i}p_c\sin{\theta_l}/\sqrt{2}$. Since $\tilde{H}_{\text{eff}, 00}=\tilde{H}_{\text{eff}, 11}=\tilde{H}_{\text{eff}, 22}$ and $\tilde{H}_{\text{eff}, 01}=\tilde{H}_{\text{eff}, 12}=\tilde{H}_{\text{eff}, 02}^{*}$, thus due to hermiticity we have to list only two matrix elements:
\begin{dmath}
    \tilde{H}_{\text{eff}, 00}= \langle 0\lvert\hat{\tilde{H}}_{\text{eff}}\rvert 0\rangle=\frac{1}{30} \left(-57 \sqrt{3} \Gamma_0+\frac{72 \sqrt{3} \Gamma_0^2 \left(324 \Gamma_0^3+45 \Gamma_0 \eta ^2
   p_c^2 \Omega ^2-20 \eta ^3 p_c^3 \Omega ^3\right)}{324 \Gamma_0^4+25 \eta ^4 p_c^4 \Omega ^4}+5 \Omega  \left(4 \eta
   ^2+3 p_c^2\right)\right),
\end{dmath}

\begin{dmath}
    \tilde{H}_{\text{eff}, 01}= \langle 0\lvert\hat{\tilde{H}}_{\text{eff}}\rvert 1\rangle=\frac{e^{-\frac{3 p_c^2}{4}}}{24 \left(324 \Gamma_0^4+25 \eta ^4 p_c^4 \Omega ^4\right)}(-1944 \left(\sqrt{3}+3 i\right) \Gamma_0^5-648 \left(\sqrt{3}-3 i\right) \Gamma_0^3 \eta ^2 p_c^2 \Omega ^2+36 \Gamma_0^2 \eta ^2 p_c^2 \Omega ^3 \left(24 \left(1+3 i \sqrt{3}\right) \eta ^2+\left(-9-27 i \sqrt{3}\right) p_c^2-8 \left(\sqrt{3}+3
   i\right) \eta  p_c\right)+324 \left(1+i \sqrt{3}\right) \Gamma_0^4 \Omega  \left(3 p_c^2-8 \eta ^2\right)+30 \left(7
   \sqrt{3}-3 i\right) \Gamma_0 \eta ^4 p_c^4 \Omega ^4+25 \left(1+i \sqrt{3}\right) \eta ^4 p_c^4 \Omega ^5 \left(3
   p_c^2-8 \eta ^2\right))
\end{dmath}

Diagonalizing this matrix $\langle l'\lvert\hat{\tilde{H}}_{\text{eff}}\rvert l\rangle$ for $\eta>\eta_c$ we get as eigenstates exactly the states, described by Eq.~(14) from the main text.

\subsection{Non-chiral waveguide}
\begin{figure}[h] 
    \centering
    \includegraphics[width=0.6\columnwidth]{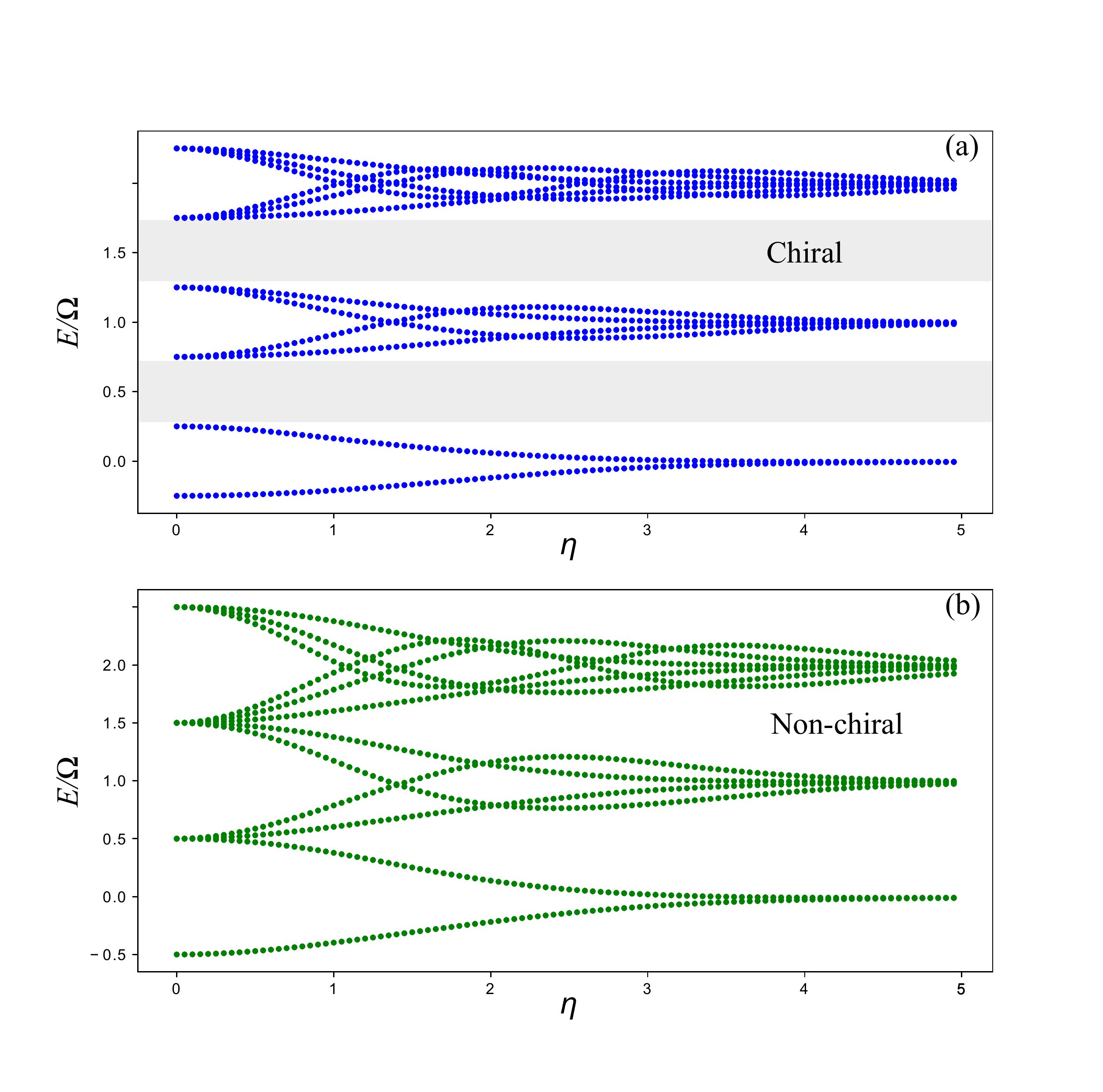}
    \caption{Comparison os the energy spectra $E(\eta) $ between (a) the perfect chiral waveguide (top panel) and (b) the non-chiral waveguide (lower panel) for the system with 2 qubits and a single photon in the system. The gaps between different groups on the dispersion branches are marked with grey rectangles in the top panel. For a non-chiral waveguide these gaps disappear. Here we assumed $\phi=\pi/(qR)$.} 
    \label{fig:chiral_vs_non_chiral}
\end{figure}

Let us consider the case, when the waveguide is non-chiral. Here we focus on the case, when there are 2 qubits suspended above the waveguide. Then the effective Hamiltonian up to the second order of the qubit-photon couplings $g_{1,2}$ reads
\begin{align}
&\hat{H}_{\rm eff}=\sum_j \omega_x\sigma_j^{+}\sigma_j+\sum_j\Omega \hat{a}_j^{\dagger}\hat{a}_j\nonumber\\-&\frac{\Gamma_0}{2}\sum_{ i<j}\left[\mathrm{i}\sigma_i^{+}\sigma_je^{\mathrm{i}qR\phi_{ij}}e^{\mathrm{i}\eta(\hat{a}_i+\hat{a}_i^{\dagger}-\hat{a}_j-\hat{a}_j^{\dagger})}+\mathrm{H.c.}\right]\nonumber\\
-&\frac{\Gamma_1}{2}\sum_{ i<j}\left[\mathrm{i}\sigma_i^{+}\sigma_je^{-\mathrm{i}qR\phi_{ij}}e^{-\mathrm{i}\eta(\hat{a}_i+\hat{a}_i^{\dagger}-\hat{a}_j-\hat{a}_j^{\dagger})}+\mathrm{H.c.}\right],  \label{H_eff_non_chiral}
\end{align}
where $q=\omega_x/v$, $\Gamma_{0,1}=g_{1,2}^2/v$ are the radiative decay rate of a single qubit into modes with opposite chirality, and $\eta=q u_0$ is the dimensionless optomechanical interaction. Below we present a plot of the comparison between the two cases: a perfect chiral waveguide with $\Gamma_0=0.5\Omega$, $\Gamma_1=0$ and a non-chiral waveguide with $\Gamma_0=\Gamma_1=0.5\Omega$. We clearly see, that the spontaneous breaking of $\mathrm{Z}_2$ symmetry is not sensible to the chirality of the waveguide. If the waveguide is nearly chiral (i.e. $\Gamma_1\ll\Gamma_0$) then the gaps between different groups of the dispersion branches $E(\eta)$ are smaller, than for a perfectly chiral waveguide (Fig.~\ref{fig:chiral_vs_non_chiral}a). For a non-chiral waveguide (i.e. $\Gamma_0=\Gamma_1$) these gaps completely vanish, as shown on the lower panel of the figure below (Fig.~\ref{fig:chiral_vs_non_chiral}b). The figure below obtained by numerical diagonaliztion of the Hamiltonian above, after projecting on the subspace with a single photon in the system.
\newpage
\section{$\mathds{Z}_n$ symmetry}

Firstly, we want to diagonalize the following matrix, which is included in hamiltonian $\Phi H \Phi^\dagger$ in the term with $\Gamma_0$:
\begin{align}
	F = \sum_{i > j} \left( - \mathrm{i} |{i}\rangle \langle{j}| \right) + \mathrm{H.c.}
\end{align}

Let us introduce matrix $M$:
\begin{align}
	M =
	\begin{pmatrix}
		0 & 0 & \cdots &  & 0 & -1\\
		1 & 0 &  &  &   & 0\\
		0 & \ddots & \ddots & \ddots & & \vdots\\
		\vdots & \ddots & \\
		& & & & & 0\\
		0 & \cdots &  & 0 & 1 & 0
	\end{pmatrix}
\end{align}
$M^k$ has only two non-zero diagonals: all elements of $(-k)$-th diagonal are equal to 1, elements on $(n - k)$-th diagonal are equal to $-1$. Thus, $F$ may be written in powers of $M$:
\begin{align}
	F = -\mathrm{i} \sum_{k=1}^{n-1} M^k,
\end{align}
therefore, $F$ commutes with $M^k$ $\forall\, k \in \mathds{N} \cup \{0\}$. We now define new operator $L = e^{\mathrm{i} \pi / n} M$ ($L^n = 1$) which generates the following cyclic group of order $n$:
\begin{align}
	\mathds{Z}_n:\ \left\{1, L, L^2, \ldots, L^{n-1} \right\}. \label{eq:Zn1}
\end{align}
Matrix $F$ commutes with all elements of this group and, consequently, possesses $\mathds{Z}_n$ symmetry.

We now want to find what irreducible representations appear in the given representation $D$~\eqref{eq:Zn1}:
\begin{align}
	D = \underset{\alpha}{\oplus} m_\alpha D_\alpha,\quad m_\alpha = \frac{1}{n_G} \sum_{g \in G} \chi^*(g) \chi_\alpha(g).
\end{align}
where $n_G$ is the number of elements in group $G$, $\chi_\alpha (g)$ -- character of the element $g$ in irrep $\alpha$. Since in our case $\chi(g) = n_G \delta_{g, e} = n \delta_{g, e}$ and $\chi_\alpha(e) = 1$, we have  $m_\alpha = 1 \, \forall \, \alpha$.

According to Wigner's theorem eigenvectors of the matrix $F$ belong to invariant subspaces of cyclic group representation~\eqref{eq:Zn1}. It is known that projection operator on the invariant subspace corresponding to irreducible representation $\alpha$ is given by the following formula:
\begin{align}
	P^{(\alpha)} = \frac{n_{\alpha}}{n_G} \sum_{g \in G} \chi_\alpha^*(g) D(g),
\end{align}
where $n_\alpha$ is the dimension of the irrep and $D(g)$ is a matrix of element $g$ in the given representation. All irreducible representations of a cyclic group are one-dimensional, so all invariant subspace are also one-dimensional, and it is possible to obtain eigenvector of the matrix $F$ from the structure of each projector $P_m$, $m=1, \ldots, n$:
\begin{align}
	P_m = \frac{1}{\sqrt{n}} \sum_{k=1}^{n} \exp \left[-\mathrm{i}\frac{2\pi (m-1)(k-1)}{n} \right] \exp\left[ \mathrm{i} \frac{\pi (k-1)}{n} \right] M^{k-1} = \frac{1}{\sqrt{n}} \sum_{k=1}^n \exp \left[ \mathrm{i} \frac{2 \pi (3/2 - m) (k-1)}{n} \right] M^{k-1},
\end{align}
please note that normalization factor was changed for convenience.
Let us write $P_m$ as a set of vector-columns:
\begin{align}
	P_m = \left(\mathbf{p}_1^{(m)},\, \mathbf{p}_2^{(m)},\, \mathbf{p}_3^{(m)},\,\ldots, \mathbf{p}_n^{(m)} \right).
\end{align}
If $\mathbf{p}_k^{(m)} \neq 0$ then $\mathbf{p}_k^{(m)}$ -- eigenvector of the matrix $F$ because all subspaces are one-dimensional. Due to the structure of $P_m$ all $\mathbf{p}_k^{(m)}$ are not equal to zero-vector, so that $\mathbf{p}_1^{(m)}$ is an eigenvector of $F$.

Finally, unitary transformation diagonalizing matrix $F$ (as $U^\dagger F U$) is
\begin{align}
	U = \left( \mathbf{p}_1^{(1)},\, \mathbf{p}_1^{(2)},\, \ldots,\, \mathbf{p}_1^{(n)} \right),
\end{align}
and its matrix elements are given as follows
\begin{align}
	U_{ij} = \frac{1}{\sqrt{n}} \exp \left[ \frac{2 \pi \mathrm{i} (3/2 - j) (i - 1)}{n} \right].
\end{align}

Let us find how an operator
\begin{align}
	A_{ij} = \mathrm{i} (a_j - a_j^\dagger) \delta_{ij},
\end{align}
which is proportional to the part of the Hamiltonian~\eqref{eq:H_after_Phi}, is transformed by the obtained unitary matrix $U$:
\begin{align}
	\begin{aligned}
		\left( U^\dagger A U \right)_{ij} &= \sum_{k,l} \left( U^\dagger \right)_{ik} A_{kl} U_{lj} = \sum_{k,l} U_{ki}^* U_{lj} A_{kl} = \sum_{k,l} \mathrm{i} U_{ki}^* U_{lj} (a_k - a_k^\dagger) \delta_{kl} = \sum_{k} \mathrm{i} U_{ki}^* U_{kj} (a_k - a_k^\dagger) = \\
		& = \frac{\mathrm{i}}{n} \sum_{k=1}^n \exp \left[ \frac{2\pi \mathrm{i}(i - j)(k - 1)}{n} \right] (a_k - a_k^\dagger).
	\end{aligned}
\end{align}
Defining new bosonic operators,
\begin{align}
	b_i = \sum_{j=1}^n V_{ij} a_j,\quad V_{ij} = \frac{1}{\sqrt{n}} \exp \left[ \frac{2\pi \mathrm{i} (i-1)(j-1)}{n} \right],\quad V^\dagger V = 1,
\end{align}
leads to the simple expressions for matrix elements of $U^\dagger A U$:
\begin{align}
	\begin{gathered}
		\left( U^\dagger A U \right)_{ii} = \frac{\mathrm{i}}{\sqrt{n}} (b_1 - b_1^\dagger),\\
		\left( U^\dagger A U \right)_{ij} = \frac{\mathrm{i}}{\sqrt{n}} (b_{i - j + 1} - b_{n - i + j + 1}^\dagger),\quad i>j,\\
		\left( U^\dagger A U \right)_{ij} = \frac{\mathrm{i}}{\sqrt{n}} (b_{n + i - j + 1} - b_{j - i + 1}^\dagger),\quad i < j.
	\end{gathered}
\end{align}
After shifting a bosonic operator of the center of mass as $b = b_1 + \mathrm{i} \eta/\sqrt{n}$, this mode decoples and the hamiltonian transformed by $U$ has only one non-diagonal matrix which is proportional to
\begin{align}
	B = U^\dagger A U - \mathrm{i}(b_1 - b_1^\dagger). \label{eq:B_matrix}
\end{align}

Let $R$ be the folowing operator
\begin{align}
	R = \exp \left[ \frac{2\pi \mathrm{i}}{n} \left( \sum_{k=1}^{ \lfloor(n-1)/2\rfloor} k\left(b_{k+1}^\dagger b_{k+1} -  b_{n-k+1}^\dagger b_{n-k+1} \right) + \frac{n}{2} \delta_{n\, \mathrm{mod}\, 2,\, 0} b_{n/2+1}^\dagger b_{n/2+1} \right) \right] \Lambda = \exp (S_b) \Lambda, \label{eq:R_operator}
\end{align}
where $\Lambda$ is $n \times n$ matrix: $\Lambda_{ij} = \delta_{ij} e^{2\pi \mathrm{i}(j-1)/n}$. $R$ generates cyclic group:
\begin{align}
	\mathds{Z}_n{:}\ \{1, R, R^2,\ldots, R^{n-1}\}.
\end{align}
Every diagonal parts of the Hamiltonian $U^\dagger \Phi H \Phi^\dagger U$ commutes with all integer powers of $R$, and it is shown that the same statement is true for operator $B$.
\begin{align}
	\comm{R}{B}_{ij} = \sum_{l=1}^n \left( R_{il}B_{lj} - B_{il}R_{lj} \right) = \sum_{l=1}^n \left( \delta_{il} \lambda_i \exp(S_b)B_{lj} - B_{il} \delta_{lj} \lambda_j \exp(S_b) \right) = \lambda_i \exp(S_b) B_{ij} - \lambda_j B_{ij} \exp(S_b)
\end{align}
According to the definitions~\eqref{eq:B_matrix} and~\eqref{eq:R_operator},
\begin{align}
	\exp(S_b) B_{ij} = \exp \left[ \frac{2\pi \mathrm{i}}{n} (j-i) \right] B_{ij} \exp(S_b) = \beta_{ij} B_{ij} \exp(S_b)
\end{align}
with $\beta_{ij} = e^{2\pi \mathrm{i} (j-i)/n}$.
Therefore,
\begin{align}
	\begin{aligned}
		\comm{R}{B}_{ij} & = \lambda_i \left[ \exp(S_b) B_{ij} - \frac{\lambda_j}{\lambda_i} B_{ij} \exp(S_b) \right] = \lambda_i \left[ \beta_{ij} - \frac{\lambda_j}{\lambda_i} \right] B_{ij} \exp(S_b) = \\
		& = \lambda_i \left[ e^{2\pi \mathrm{i} (j-i)/n} - e^{2\pi \mathrm{i} (j-i)/n} \right] B_{ij} \exp(S_b) = 0.
	\end{aligned}
\end{align}
The same commutation relation is true for all integer powers of $R$:
\begin{align}
	\comm{R^m}{B}_{ij} = \lambda_i^m \exp(mS_b)B_{ij} - \lambda_j^m B_{ij} \exp(mS_b) = \lambda_i^m \left[ \beta_{ij}^m - \left( \frac{\lambda_j}{\lambda_i} \right)^m \right] \exp(mS_b) = 0.
\end{align}
Thus, the Hamiltonian possesses global $\mathds{Z}_n$-symmetry.

\end{document}